\documentclass[10pt,twocolumn,showpacs,preprintnumbers,%
               aps,prd]{revtex4-1}
\usepackage[dvips]{graphicx}
\usepackage{amsmath,amssymb}
\usepackage{color}

\newcommand{\utwo}{\mathrm{U(2)\times U(2)}}
\newcommand{\Nf}{N_{\text{f}}}
\newcommand{\UA}{\mathrm{U(1)_A}}

\begin{document}

\title{Second-order and Fluctuation-induced First-order
       Phase Transitions\\
       with Functional Renormalization Group Equations}

\author{Kenji Fukushima}    
\affiliation{Department of Physics, Keio University,
             Kanagawa 223-8522, Japan}
\author{Kazuhiko Kamikado}  
\affiliation{Yukawa Institute for Theoretical Physics,
             Kyoto University, Kyoto 606-8502, Japan}

\author{Bertram Klein}      
\affiliation{Technische Universit\"at M\"unchen,
             James-Franck-Strasse 1, 85748 Garching, Germany}

\begin{abstract}
 We investigate phase transitions in scalar field theories using the
 functional renormalization group (RG) equation.  We analyze a system
 with $\utwo$ symmetry, in which there is a parameter $\lambda_2$ that
 controls the strength of the first-order phase transition driven by
 fluctuations.  In the limit of $\lambda_2\to0$, the $\utwo$ theory is
 reduced to an $\mathrm{O}(8)$ scalar theory that exhibits a
 second-order phase transition in three dimensions.  We develop a new
 insight for the understanding of the fluctuation-induced first-order
 phase transition as a smooth continuation from the standard RG flow
 in the $\mathrm{O}(8)$ system.  In our view from the RG flow diagram
 on coupling parameter space, the region that favors the first-order
 transition emerges from the unphysical region to the physical one as
 $\lambda_2$ increases from zero.  We give this interpretation based
 on the Taylor expansion of the functional RG equations up to the
 fourth order in terms of the field, which encompasses the
 $\varepsilon$-expansion results.  We compare results from the
 expansion and from the full numerical calculation and find that the
 fourth-order expansion is only of qualitative use and that the
 sixth-order expansion improves the quantitative agreement.
\end{abstract}
\preprint{YITP-10-66}
\maketitle


\section{INTRODUCTION}

Phase transitions occur in many different physical systems.  In this
paper we shall address a fluctuation-induced first-order phase
transition or the Coleman-Weinberg potential
\cite{Coleman:1973jx,Halperin:1973jh,Bak:1976zza,Paterson:1980fc,%
Pisarski:1980ix,Iacobson:1981jm,Pisarski:1983ms,Bornholdt:1994rf,%
Alford:1993br,Litim:1994jd,Litim:1996nw} in the system with $\utwo$
symmetry, as well as a second-order one in the system with
$\mathrm{O}(8)$ symmetry
\cite{Baym:1977qb,Kogut:1981ez,AmelinoCamelia:1992nc,Chiku:1998kd,%
Roder:2005vt,Cooper:2005vw,Le_Guillou:1977ju,Schaefer:1999em,%
Bohr:2000gp,Shafer:2003ym,Blaizot:2006rj,ParisenToldin:2003hq,%
Braun:2007td} as a particular limit of the $\utwo$ theory.  This
particular choice of the global symmetries is motivated by those
relevant for phase transitions in the strong interactions.

The fundamental theory of the strong interactions is Quantum
Chromodynamics (QCD), which describes the dynamics of quarks and
gluons.  It is known that at a temperature $T$ of the same order as
the typical QCD scale, $\Lambda_{\text{QCD}}$, QCD undergoes two types
of phase transitions --- the chiral phase transition and the quark
deconfinement transition \cite{review}.  In particular at the chiral
phase transition the behavior of the system should be characterized by
global chiral symmetry \cite{Pisarski:1983ms,Wilczek:1992sf}.

For $\Nf$ flavors of massless quarks the QCD Lagrangian is invariant
under a flavor rotation of the
$\mathrm{U}(\Nf)_{\text{L}}\times \mathrm{U}(\Nf)_{\text{R}}$ symmetry
for left- and right-handed quarks.  Because of the axial anomaly, the
subgroup $\UA$ is explicitly broken (down to $\mathrm{Z}_{2\Nf}$) on
the quantum level, so that in this case the (continuous) symmetry is
reduced to $\mathrm{SU}(\Nf)_{\text{L}} \times \mathrm{SU}(\Nf)_{\text{R}}
 \times \mathrm{U}(1)_{\text{V}}$.  Here, the vector symmetry
$\mathrm{U}(1)_{\text{V}}$ corresponds to the conserved baryon charge
and is not broken in the normal phase.

In the QCD vacuum chiral symmetry is spontaneously broken by a nonzero
chiral condensate \cite{Nambu:1961tp} according to
$\mathrm{SU}(\Nf)_{\text{L}} \times \mathrm{SU}(\Nf)_{\text{R}}
 \to \mathrm{SU}(\Nf)_{\text{V}}$.  In the two-flavor case the
symmetry breaking pattern,
$\mathrm{SU(2)_L}\times \mathrm{SU(2)_R} \to \mathrm{SU(2)_V}$, is
equivalent to $\mathrm{SO}(4) \to \mathrm{SO}(3)$, and hence can be
mapped onto a (scalar) meson model with $\mathrm{O}(4)$ symmetry
\cite{GellMann:1960np}.  This provides a motivation to investigate
phase transitions of the $\mathrm{O}(N)$ scalar theory in the
finite-$T$ formalism in $d=4$ dimensions, or in the zero-$T$ formalism
in $d=3$ dimensions, if we assume that the critical behavior is
described with a dimensionally reduced theory
\cite{Appelquist:1981vg}.  In this paper we address the
Renormalization Group (RG) flow of the theory in the dimensionally
reduced description only.  This can be justified as follows:  At
finite temperature, the Euclidean time direction in a $d$-dimensional
path-integral description becomes compactified.  Because of this
finite extent in the Euclidean time direction, in a region close
enough to a critical point, the dominant long-range fluctuations will
be unable to propagate in the time direction and the system will in
effect become $(d-1)$ dimensional.  For a finite-$T$ theory in $d=4$
space-time it is therefore expected that the critical behavior can be
described in terms of an effective $d=3$ dimensional theory at $T=0$
with its couplings (which are $T$-dependent) set close to their
critical values.

In contrast to the situation described above, if the axial symmetry is
\textit{effectively} restored close to the transition temperature
(e.g.\ because of the instanton suppression at high $T$
\cite{Gross:1980br,Shuryak:1993ee,Schafer:2002ty}), the relevant
symmetry for the $\Nf=2$ transition is then
$\mathrm{U(2)_L}\times \mathrm{U(2)_R}$, which leads us to the
analysis on the $\utwo$ scalar theory.

Assuming that a particular phase transition is of second order and
displays critical behavior, one can rely on universality arguments to
describe the system in terms of an effective Landau-Ginzburg-Wilson
functional which is determined solely from the symmetries and the
dimensionality of the system.  The description in terms of such an
effective theory is often much simpler than the original microscopic
theories, but still captures the long-range phenomena at the critical
point.

In the investigation of critical phenomena, the correct treatment of
low-momentum modes is of paramount importance.  Close to a critical
point, physical properties of a system are dominated by these
low-lying soft modes.  Since the system has no intrinsic length scale
at the critical point, phenomena on all length scales contribute and
non-trivial anomalous dimensions can appear.  In the case of scalar
theories, the anomalous dimension at the critical point is generally
small, which can justify a simplified treatment.

Renormalization group methods have been used to deal with the infrared
(IR) divergences which appear in the vicinity of critical points
\cite{Wilson:1971bg}.  Historically speaking, the
$\varepsilon$-expansion around $d=4$ dimensions for theories with
upper critical dimension $d=4$ has been of great importance
\cite{Wilson:1971dc,Wilson:1973jj}.  While the RG methods are able to
deal with IR divergences, calculations of this type are still
perturbative and rely on an additional expansion in terms of small
coupling constants, which can be justified by the fact that the
coupling constants are of order of $\varepsilon$ in the vicinity of
the fixed points.  There is a countless number of works on the
application of RG methods to the second-order phase transition in the
$\mathrm{O}(N)$ theory
\cite{Pisarski:1983ms,Le_Guillou:1977ju,Schaefer:1999em,Bohr:2000gp,%
Shafer:2003ym,Blaizot:2006rj,ParisenToldin:2003hq,Braun:2007td}.

On the other hand, the theory with $\utwo$-symmetry requires more
investigation.  In the context of the chiral phase transition in QCD,
such an effective theory has been extensively discussed by Pisarski
and Wilczek \cite{Pisarski:1980ix,Pisarski:1983ms,Wilczek:1992sf}.  In
their results the $\varepsilon$-expansion analysis does not find an IR
stable fixed point which could control a continuous transition.  This
indicates that the transition is presumably driven by fluctuations to
be of first order.  [See also
\cite{Bak:1976zza,Alford:1993br,Yamagishi:1981qq}].

This question of the order of the chiral phase transition including
the possibility of the effective $\UA$ symmetry restoration has been
further analyzed in and beyond the $\varepsilon$-expansion in e.g.\
\cite{Calabrese:2002bm,Butti:2003nu,Vicari:2007ma} using
field-theoretical methods and in
\cite{Jungnickel:1995fp,Berges:1995mw,Berges:1996ja,Berges:1996ib}
using functional RG
methods \cite{Wetterich:1992yh,Berges:1998nf,Berges:2000ew} as well as
the finite-$T$ lattice-QCD simulation \cite{lattice}.  Closely related
first-order phase-transitions have been investigated in
\cite{Litim:1994jd,Litim:1996nw,Alford:1993br,Bergerhoff:1995zq} with
the functional RG.\ \ In a gauge theory setting, the functional RG has
been used to estimate effects of the anomaly on chiral symmetry
breaking \cite{Gies:2002hq} and on the phase transition line
\cite{Braun:2008pi} in QCD.\ \ Also, anomaly effects on the mass
spectrum have been described in \cite{Alkofer:2008et} by means of
non-perturbative Dyson-Schwinger equations.

The functional RG approach does not rely on an expansion in terms of a
small coupling and is capable of describing the full effective
potential.  This enables us to directly analyze the critical behavior
associated with not only the second-order but also first-order phase
transitions.  For reviews of the method see e.g.\
\cite{Berges:1998nf,Litim:1998nf,Bagnuls:2000ae,Berges:2000ew,%
Polonyi:2001se,Delamotte:2003dw,Pawlowski:2005xe,Gies:2006wv,%
Delamotte:2007pf,Sonoda:2007av,Rosten:2010vm}.

We use a functional RG equation to describe the behavior of the
$\utwo$ matrix model at the fluctuation-induced first-order phase
transition in the same way as done by Berges and Wetterich
\cite{Berges:1996ja}.  We establish the connection between the
functional RG flow equations in the formulation according to
\cite{Wetterich:1992yh} (i.e. the Wetterich equation) and the
$\varepsilon$-expansion.  While there is some literature on the
connection between the functional RG and perturbative calculations
\cite{Alford:1993br,Blaizot:2006rj,Papenbrock:1994kf,Litim:2002xm},
this connection we address is a useful starting point to explore the
general RG flow diagram.  Moreover, to our knowledge, nobody has
explicitly retrieved the $\varepsilon$-expansion results from the
Wetterich equation for the $\utwo$ system.  [See also
\cite{Yamagishi:1981qq,Shen:1993yx}.  The connection of the
$\varepsilon$-expansion to functional RG equations in a related
formalism and the emergence of the Wilson-Fisher fixed-point in this
approach have been pointed out in \cite{Rosten:2010vm}.]

Since we are motivated by an interest in the QCD phase transition at 
finite temperature, we will adhere to the QCD terminology throughout
this paper.  That is, we will refer to the $8$-component fields in the
$\mathrm{O}(8)$ and $\utwo$ theory as the $\sigma$-field for the
condensing mode which acquires a non-zero expectation value and as the
$\vec{\pi}$-fields for the $3$ pseudo-scalar components that are
Nambu-Goldstone bosons.  These $4$ mesons have parity partners,
namely, $\eta$-field and $\vec{a}_0$-fields.  Apart from this,
nevertheless, our treatment is generic and is completely parallel to
the earlier works by Berges and Wetterich~\cite{Berges:1996ja} and by
Berges, Tetradis, and Wetterich~\cite{Berges:1996ib}.  In the next
section, as a convenience to readers, we will briefly summarize our
aims in this paper and discuss where they advance beyond those earlier
works.

We intend the current calculation as a starting point for further
investigations of the transition in the framework of more realistic
phenomenological models for QCD thermodynamics
\cite{Fukushima:2003fw,Ratti:2005jh,Schaefer:2007pw,Fu:2007xc,%
Ciminale:2007sr,Schaefer:2004en,Nakano:2009ps,Braun:2010vd,%
Sano:2009wd}.  In such models gauge degrees of freedom can also be
taken into account, but the possibility of a fluctuation-induced
change of the order of the transition has not yet been fully
investigated due to technical complexity.  The inclusion of such
effects beyond the mean-field level in these models would be a very
interesting problem in the future.

We will organize this paper as follows:  We first present a summary of
our central results in Sec.~\ref{sec:results}.  Then we make a quick
review over the functional RG formalism in Sec.~\ref{sec:formalism}.
We discuss the analytical and numerical calculations in great details
in Sec.~\ref{sec:u2u2}.  Section~\ref{sec:conclusions} contains our
concluding remarks.


\section{CENTRAL RESULTS}
\label{sec:results}

The phase transitions of the $\utwo$ scalar theory have been
extensively studied in earlier works in the functional RG formalism
\cite{Berges:1996ja,Berges:1996ib}.  Although there are minor
differences in the technical setup, we have performed numerical
calculations at the same level of approximation and truncation as used
by Berges, Tetradis, and Wetterich in
\cite{Berges:1996ja,Berges:1996ib}.  Before we discuss the details of
the calculation, we therefore wish to point out the qualitatively new
aspects of this work.

First of all, we would like to emphasize that we use a very simple
form of the flow equations (given by Eqs.~\eqref{eq:FRG_U} and
\eqref{eq:FRG_V} below) in the local potential approximation which
does no longer involve momentum integrations.  This difference from
the equations in \cite{Berges:1996ja,Berges:1996ib} arises from a
different choice of the IR regulator function.  Apart from the
wave-function renormalization, which we neglect, the information
contained in the RG flow equation is equivalent.  Since the content is
equivalent, the simpler expressions are advantageous, both for an
analysis of the structure of the RG flow and for numerical
calculations.  In fact, as we will demonstrate explicitly later in the
discussion, we can reproduce the result from the
$\varepsilon$-expansion in a short calculation of only a few lines and
make the connection between the functional RG equation and the
perturbative method very transparent.  This is certainly not a
qualitative breakthrough, but such a technical improvement provides
useful progress for practical calculations.

\begin{figure}
 \includegraphics[width=0.97\columnwidth]{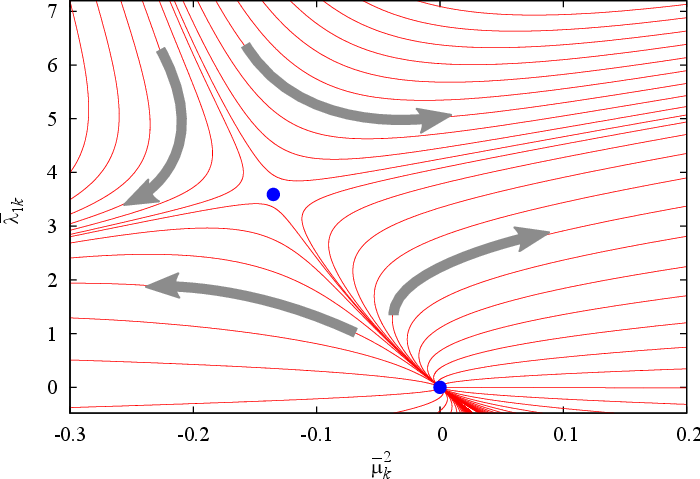}
 \caption{RG flow diagram in the $\bar{\mu}_k^2$-$\bar{\lambda}_{1k}$
   plane for the $d=3$ $\mathrm{O}(8)$ scalar theory that is realized
   from the $\utwo$ theory with $\bar{\lambda}_{2\Lambda}=0$ chosen
   initially.  The arrows indicate the direction from larger to
   smaller $k$ and the dots represents the fixed points.}
 \label{fig:o8}
\end{figure}

Second, in this work we propose a new point of view for understanding
the fluctuation-induced first-order phase transition.  If the phase
transition is of second order, the flow diagram and the fixed-point
structure can be studied very well in the space of the coupling
parameters.  For instance, in the case of the $\mathrm{O}(N)$ scalar
theory in three dimensions with a field
$\phi=(\phi_1, \ldots \phi_N)^T$, the relevant operators in terms
of the squared field $\varphi=\phi^2$ are
$\varphi$ and $\varphi^2$.  The (IR-scale dependent) effective
potential can thus be written as
\begin{equation}
 U_k(\varphi) = \frac{1}{2}\mu_k^2 \varphi
  + \frac{1}{4!}\lambda_{1k}\varphi^2 .
\label{eq:U_o8}
\end{equation}
The relevant coupling parameters, $\mu_k^2$ and $\lambda_{1k}$, run as
the IR scale $k$ becomes smaller and fluctuations are integrated out.
Using dimensionless variables $\bar{\mu}_k^2\equiv k^{-2}\mu_k^2$ and
$\bar{\lambda}_{1k}=k^{-1}\lambda_{1k}$, we obtain Fig.~\ref{fig:o8}.
Starting from a set of differential flow equations to be derived
later, the figure shows how these dimensionless couplings change under
a change of the scale $k$.


\begin{figure}
 \includegraphics[width=0.96\columnwidth]{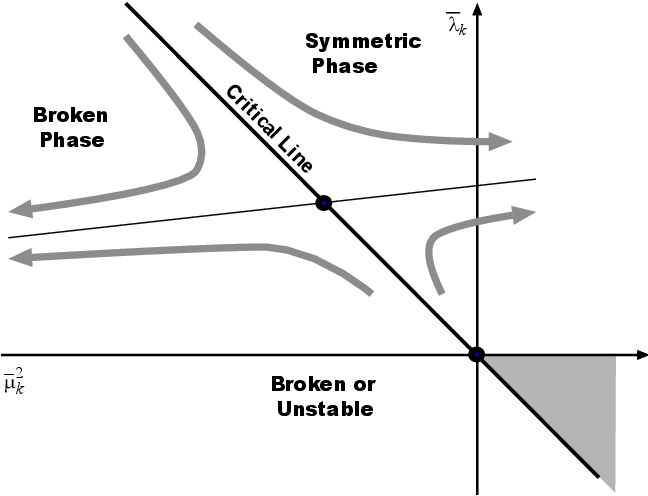}
 \caption{General structure of the RG flow for the $d=3$
   $\mathrm{O}(N)$ scalar theory.  In the shaded region the flow goes
   to $\bar{\mu}_0^2>0$ and $\bar{\lambda}_{10}<0$ that is the
   ``necessary condition'' for a double-well shape in the effective
   potential~\eqref{eq:U_o8}.}
 \label{fig:schematic}
\end{figure}

The interpretation of such a RG-flow diagram is merely textbook
knowledge.  We sketch an illustration in Fig.~\ref{fig:schematic} to
extract the essential features from Fig.~\ref{fig:o8}.  If the flow
goes toward the left side to smaller values of $\bar{\mu}_k^2$, the
symmetric state at $\varphi=0$ becomes more and more unstable and the
system falls into the symmetry-broken phase with $\varphi\neq0$.  If
the flow goes in the opposite direction toward the right to larger
values of $\bar{\mu}_k^2$, on the other hand, the symmetric phase is a
stable ground state.  Therefore there is a critical line separating
these two regimes.  If the flow starts on the critical line, the
system approaches the IR fixed-point which is the critical point
characterizing the phase transition.


\begin{figure}
 \includegraphics[width=0.6\columnwidth]{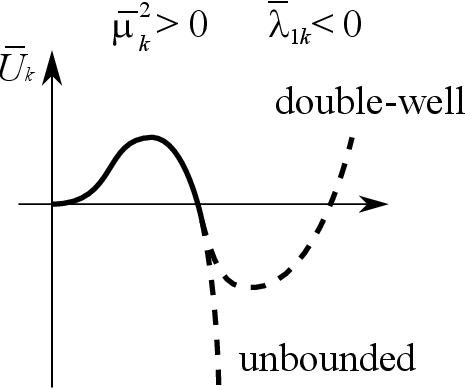}
 \caption{Potential shape for $\bar{\mu}_k^2>0$ and
   $\bar{\lambda}_{1k}<0$.  Depending on the sixth-order coefficient
   the potential may take a double-well form.}
 \label{fig:potential}
\end{figure}

In the shaded region in Fig.~\ref{fig:schematic} the flow goes deeper
into the region with $\bar{\mu}_0^2>0$ and $\bar{\lambda}_{10}<0$.
Usually such flow patterns are physically meaningless in the analysis
of the $\mathrm{O}(N)$ theory because negative ${\lambda}_{10}$ makes
the potential~\eqref{eq:U_o8} unbounded and the theory is not well
defined there.  Nevertheless, from the interest in the first-order
transition, this shaded region is interesting.  This is because, in
view of the potential form~\eqref{eq:U_o8}, the destination of the
flow, $\bar{\mu}_0^2>0$ and $\bar{\lambda}_{10}<0$, is just the
``necessary condition'' to realize a double-well shape (see
Fig.~\ref{fig:potential}).  Of course this is not a ``sufficient
condition'' and the existence of the double-well shape depends on how
large the sixth-order coupling constant $\zeta_1$ associated with
$\varphi^3$ is (see Eq.~\eqref{eq:expanded}).  If $\zeta_1$ is
too small, the potential is not stabilized until very large values of
$\varphi$ have been reached, for which an expansion like
Eq.~\eqref{eq:U_o8} is unsuitable.  If $\zeta_1$ is too large, the
second minimum at $\varphi\neq0$ simply vanishes and the symmetric
phase is always the most stable.  Nevertheless, we can say that, if
there is a region where the first-order phase transition exists, the
flow must satisfy the necessary condition, meaning that the flow must
be headed for the region with $\bar{\mu}_0^2>0$ and
$\bar{\lambda}_{10}<0$.


The $\utwo$ theory has at least one more coupling constant $\lambda_2$
in addition to $\mu^2$ and $\lambda_1$ (and $\zeta_1$) as defined
in Sec.~\ref{sec:u2u2}.  For $\lambda_2=0$ the theory is reduced into
the $\mathrm{O}(8)$ scalar theory.  Then, recalling the discussions in
the previous paragraph, we can formulate our expectations about the
evolution of the phase diagram.

An interesting question is the following:  How does the RG-flow
diagram or more specifically the shaded region in
Fig.~\ref{fig:schematic} change in the $\utwo$ model as the additional
coupling $\lambda_2$ increases from zero?  This question does not seem
to have been answered explicitly in the earlier works by Berges,
Tetradis, and Wetterich~\cite{Berges:1996ja,Berges:1996ib} nor by
Alford and March-Russell~\cite{Alford:1993br}, though those works
contain extensive and detailed discussions.

Answering this question is quite intriguing because, as we have
explained above, the shaded region has a direct connection to the
emergence of the double-well shape in the effective potential.  We can
understand the results in a way parallel to the $\mathrm{O}(8)$ case,
using the phase diagram in space spanned by $\bar{\mu}_k^2$ and
$\bar{\lambda}_{1k}$.  Primarily this is just a replacement of the
variables used to draw the phase diagram, but it also provides a
useful interpretation in terms of the shape of the potential and we
believe that it is worth taking a closer look at this interpretation
in the following.

\begin{figure}
 \includegraphics[width=\columnwidth]{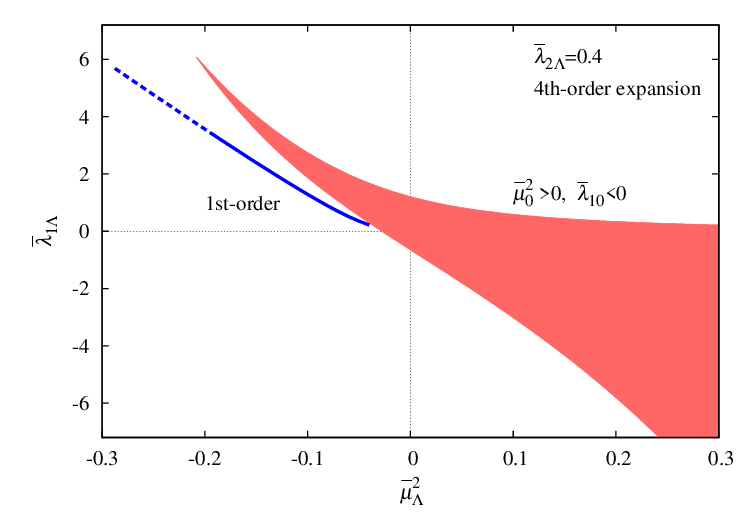}
 \caption{Initial condition region at the UV scale necessary for the
   double-well shape in the effective potential at the IR scale.
   Results are from the fourth-order Taylor expansion.  The line
   labelled with ``1st-order'' represents the phase transition points
   of fluctuation-induced first order as a result of full numerical
   evaluation in the grid method.}
  \label{fig:region04-4}
\end{figure}

\begin{figure}
 \includegraphics[width=\columnwidth]{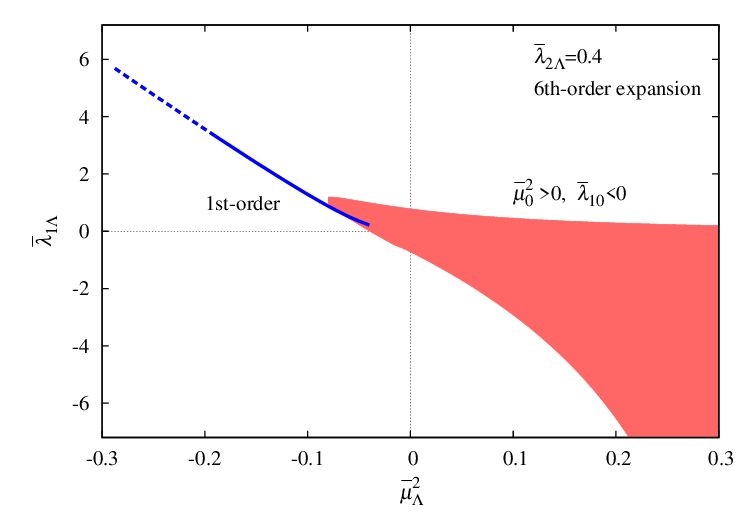}
 \caption{The shaded region indicates the initial conditions at the UV
   scale necessary to obtain the double-well shape in the effective
   potential at the IR scale.  Results are from the sixth-order Taylor
   expansion for $\bar{\lambda}_{2\Lambda}=0.4$.}
  \label{fig:region04-6}
\end{figure}

\begin{figure}
 \includegraphics[width=\columnwidth]{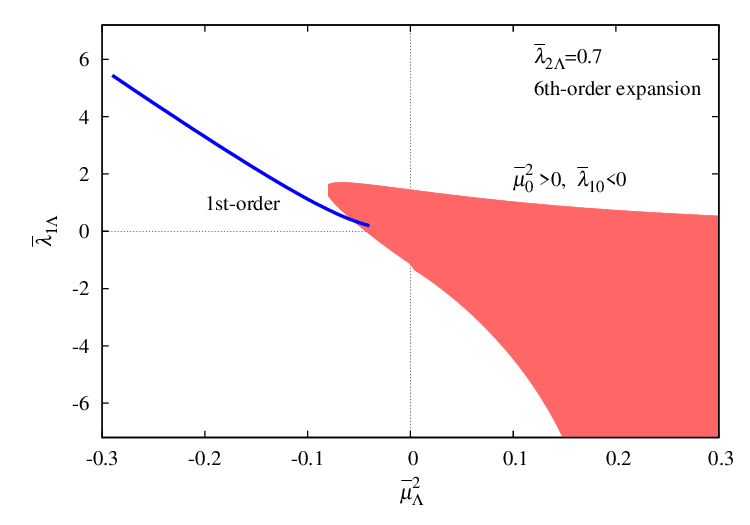}
 \caption{The shaded region indicates the initial conditions at the UV
   scale necessary to obtain the double-well shape in the effective
   potential at the IR scale.  Results are from the sixth-order Taylor
   expansion for $\bar{\lambda}_{2\Lambda}=0.7$.}
  \label{fig:region07-6}
\end{figure}

Before we discuss the results shown in Figs.~\ref{fig:region04-4},
\ref{fig:region04-6}, and \ref{fig:region07-6}, we remark that the
horizontal and vertical axes are not, strictly speaking, identical to
those in the previous Fig.~\ref{fig:o8}.  In the diagram we use
$\bar{\mu}_\Lambda^2$ and $\bar{\lambda}_{1\Lambda}$, which are the
coupling parameters at the UV scale $k=\Lambda$, to show the shaded
regions.  If the flow starts with initial $\bar{\mu}_\Lambda^2$ and
$\bar{\lambda}_{1\Lambda}$ inside the shaded region and the specified
$\bar{\lambda}_{2\Lambda}$ for each figure, it goes toward
$\bar{\mu}_0^2>0$ and $\bar{\lambda}_{10}<0$ at the IR scale
$k\approx 0$.  In the numerical calculation, we do not actually take
the limit of $k\to0$ but stop the flow at $k/\Lambda=0.1$.  This is a
prescription for the numerical calculation (i.e.\ the choice of $0.1$
is just arbitrary).  As discussed in
\cite{Alford:1993br,Berges:1996ja,Berges:1996ib}, in the context of a
first-order transition with bubble formation, it would make physical
sense to stop the flow at some point whose scale should be related to
the bubble thickness.

The RG-flow diagram for the $\utwo$ model similar to Fig.~\ref{fig:o8}
is available from the solution of the RG equations for the coupling
parameters, that is, the Taylor expansion coefficients.  In the same
way as for the $\mathrm{O}(N)$ model we have performed the Taylor
expansion for the $\utwo$ system to derive a set of partial
differential equations for $\bar{\mu}_k^2$, $\bar{\lambda}_{1k}$, and
$\bar{\lambda}_{2k}$.  Then, solving the differential equations
numerically, we have identified the shaded region corresponding to
that in Fig.~\ref{fig:schematic} in the $\mathrm{O}(N)$ theory.


Figure~\ref{fig:region04-4} shows the results from the Taylor
expansion up to the fourth order in terms of the field (or the second
order in terms of $\varphi=\phi^2$).  It is clear that our expectation
is manifestly fulfilled:  The shaded region is elongated from the
unphysical region in the $\mathrm{O}(N)$ case and spreads almost along
the critical line toward the physical region at
$\bar{\mu}_\Lambda^2<0$ and $\bar{\lambda}_{1\Lambda}>0$.

As we will describe in detail in a later section, we have carried out
a full numerical calculation without resorting to an expansion by
putting the function $U_k(\varphi)$ on a grid of discretized field
values.  In this way we can confirm the double-well potential form
already reported in the earlier works by Berges, Tetradis, and
Wetterich \cite{Berges:1996ja,Berges:1996ib}.  While our method does
not surpass the earlier works in precision, we use a regulator that
leads to much simpler flow equations.

The line labelled with ``1st-order'' in Fig.~\ref{fig:region04-4}
represents the phase transition points of fluctuation-induced first
order.  The strength of the first-order transition weakens with
decreasing $\bar{\mu}_\Lambda^2$ and eventually it becomes smaller
than the resolution in the grid method.  We indicate this by changing
the solid line to a dotted one.  We can see that the agreement between
the results from the fourth-order Taylor expansion and from the full
numerical evaluation is limited to a qualitative level.  Guided by the
expectation that the agreement would be better, we have investigated
the phase diagram from the sixth-order Taylor expansion.


We show the results from the sixth-order Taylor expansion in
Fig.~\ref{fig:region04-6}.  The phase transition line seems to be
smoothly connected to the shaded region in this case.  The agreement
is, however, not as good as we expected.  [We have checked that the
  second minimum in the potential lies in the region where $\varphi$
  is sufficiently smaller than unity in all the cases, so that the
  Taylor expansion should work in principle.]  We will discuss the
quantitative comparison for the shape of the potential with and
without expansion later.

So far, we have seen the results for a fixed choice of
$\bar{\lambda}_{2\Lambda}=0.4$.  According to
\cite{Berges:1996ja,Berges:1996ib}, the first-order phase transition
is strengthened with increasing $\bar{\lambda}_{2\Lambda}$.  We can
reconfirm this observation by repeating the calculations for larger
$\bar{\lambda}_{2\Lambda}$.  Figure~\ref{fig:region07-6} shows the
results for a larger choice of $\bar{\lambda}_{2\Lambda}=0.7$.  We can
see that the shaded region in Fig.~\ref{fig:region07-6} becomes
slightly wider than that in Fig.~\ref{fig:region04-6}.  Also the
dotted line disappears since the first-order phase transition becomes
more prominent and is resolved by the grid in the region shown.


If we choose even larger values of $\bar{\lambda}_{2\Lambda}$, we find
that the shaded region is pulled not along the critical line, but
upward toward larger values of $\bar{\mu}_\Lambda^2$, while the
position of the first-order transition line hardly moves.


\section{FORMALISM}
\label{sec:formalism}

To make our discussion as self-contained as possible, we shall briefly
review the functional RG formalism and the equations we are using in
this paper.  The functional RG is an exact formulation of quantum
field theories in the form of a functional differential
equation~\cite{Wetterich:1992yh,Berges:1998nf,Berges:2000ew}.  In the
formulation we are using, the central object is the effective action
which is calculated non-perturbatively.  In the functional RG one
defines an average effective action with the IR-cutoff scale $k$.  For
a general scalar field $\chi$ whose classical counterpart is denoted
by $\phi=\langle\chi\rangle$ here, the $k$-dependent effective action
is defined by
\begin{align}
 \Gamma_k [\phi] &= -\log\biggl[ \int [d\chi]\, \exp \Bigl(
  -S[\chi] - \Delta S_k [\chi - \phi] \notag\\
 & \qquad + \int d^d x\;
  \frac{\delta\Gamma_k [\phi]}{\delta\phi(x)}\, \bigl[
  \chi(x)-\phi(x) \bigr] \Bigr) \biggr] \;,
\label{eq:effective_action}
\end{align}
where the IR-cutoff dependence is introduced by a mass-like term,
\begin{equation}
 \Delta S_k = \frac{1}{2} \int \frac{d^d q}{(2\pi)^d} \;
  R_k(q)\, \phi(-q)\,\phi(q)
\end{equation}
in momentum space.  Here we consider $d$-dimensional space-time
generally.  We note that the momentum-dependent mass $R_k(q)$ should
cut off IR modes with $q<k$.  That is, $R_k(q)$ must satisfy the
following
requirement~\cite{Wetterich:1992yh,Berges:1998nf,Berges:2000ew};  soft
modes with $q\ll k$ should be quenched by large $R_k(q)\sim k^2$,
while $R_k(q)\sim 0$ for $q\gg k$.
Equation~\eqref{eq:effective_action} together with the properties of
$R_k(q)$ leads to the boundary condition for the average effective
action;  $\Gamma_\Lambda[\phi]=S[\phi]$ where $\Lambda$ is the UV
scale where the RG flow starts, and $\Gamma_0[\phi]=\Gamma[\phi]$
where $\Gamma[\phi]$ is the full effective action defined in the
textbook manner.  In principle, the IR-cutoff function can be
arbitrary as long as it satisfies the above-mentioned properties.  In
this paper we make use of the ``optimized'' choice of $R_k(q)$
\cite{Litim:2000ci,Litim:2001up,Pawlowski:2005xe}
proposed by Litim~\cite{Litim:2001up},
\begin{equation}
 R_k(q) = (k^2 -q^2)\, \theta(k^2 -q^2) \;,
\label{eq:cut_off}
\end{equation}
where $q$ refers to a $d$-dimensional vector in Euclidean space-time
and $\theta$ is the Heaviside step function.  
The choice of a cutoff function is in general not unique, but the
optimized choice \cite{Litim:2001up} 
ensures the correct universal behavior \cite{Litim:2005us}.  (For
finite-$T$ studies it is a better choice to use the cutoff for only
spatial momenta \cite{Litim:2006ag,Braun:2003ii}.)

Armed with these definitions, we can reach the functional RG equation
(known as the Wetterich equation~\cite{Wetterich:1992yh}) by
differentiating $\Gamma_k[\phi]$ with respect to $k$, i.e.
\begin{equation}
 \frac{\partial \Gamma_k[\phi]}{\partial k} = \frac{1}{2}
  {\rm Tr} \Biggl[ \frac{\partial R_k}{\partial k}
  \Bigl( \Gamma_k^{(2)}[\phi] + R_k \Bigr)^{-1} \Biggr] \;.
\label{eq:wetterich}
\end{equation}
Here ${\rm Tr}$ means the trace over any intrinsic indices of $\phi$
as well as the trace over the functional space.  In the above
$\Gamma_k^{(2)}[\phi]$ denotes the second-order functional derivative
of $\Gamma_k[\phi]$ with respect to the field $\phi$, and
$\Gamma_k^{(2)}[\phi]+R_k$ is thus an inverse of the full propagator
including the IR-cutoff term.  From this one can acquire an intuitive
interpretation for the left-hand side of Eq.~\eqref{eq:wetterich} as a
one-loop integration attached to the cutoff derivative,
$\partial R_k/\partial k$.  Nevertheless, since the derivation assumes
no approximation, Eq.~\eqref{eq:wetterich} is ``exact'' and contains
full quantum effects.

In practice, however, solving the functional RG equation without any
truncation is as difficult as finding an exact solution of the
theory.  We should thus make an approximation.  Here we adopt the
local potential approximation (LPA) in which the effective action
$\Gamma_k[\phi]$ is assumed to be of the form,
\begin{equation}
 \Gamma_k[\phi] = \int d^d x\, \biggl( \frac{1}{2}
  \partial_\mu \phi\, \partial_\mu \phi
  + U_k(\phi) \biggl) \;,
\label{eq:LPA}
\end{equation}
in $d$-dimensional Euclidean space-time.  The $k$-dependent effective
potential $U_k(\phi)$ is a local function of the fields.  This
approach is frequently taken in theories involving only scalar fields
and is sufficient as long as the anomalous dimension remains small.
Thanks to the particular choice~\eqref{eq:cut_off} one can easily
perform the $q$-integration after Eq.~\eqref{eq:LPA} is substituted
for $\Gamma_k[\phi]$ in Eq.~\eqref{eq:wetterich} (and $\phi$ is
assumed to be spatially constant).  
Finally the functional RG
equation simply reads
\begin{equation}
 \frac{\partial U_k}{\partial k} = K_d\, k^{d+1} \sum_i
  \frac{1}{E^2_i} \;.
\label{eq:FRG}
\end{equation}
The index $i$ refers to the field components $\phi_i$.  Here, we have
defined $K_d=S_d/(2\pi)^d$ with the volume of a $d$-dimensional sphere
$S_d$ given by
\begin{equation}
 S_d = \frac{2 \pi^{d/2}}{d\,\Gamma(d/2)} \;,
\end{equation}
that is, for example, for $d=3$ we need
\begin{equation}
 K_3 = \frac{1}{6 \pi^2} \;.
\end{equation}
The energy in the denominator of Eq.~\eqref{eq:FRG} is
\begin{equation}
 E_i^2 = k^2 + M^2_i \;,
\end{equation}
where $M_i$ denotes the mass eigenvalues of the second-derivative
matrices of the effective potential,
\begin{equation}
 M_{ij}
  = \frac{\partial^2 U_k(\phi)}{\partial\phi_i \, \partial\phi_j} \;.
\end{equation}
Although Eq.~\eqref{eq:FRG} is very simple, it
encompasses rich contents of the critical phenomena, as we will see
later.


\section{$\utwo$ SCALAR THEORY}
\label{sec:u2u2}

We are now ready to proceed to our main subject, i.e.\ the
fluctuation-induced first-order phase transition.  For this purpose we
adapt the $\utwo$ scalar theory in $d=3$ dimensions and introduce some
notations according to hadron physics.


\subsection{Notations}

We can write the Lagrangian density of the $\utwo$ theory conveniently
as \cite{Pisarski:1983ms}
\begin{equation}
 \begin{split}
 \mathcal{L} &=\frac{1}{2} \mathrm{tr}
  \bigl[ \partial_\mu\Phi \, \partial^\mu\Phi^\dagger \bigr]
  + \frac{1}{2}\mu^2 \mathrm{tr}\bigl[ \Phi \Phi^\dagger \bigr] \\
 &\qquad - g_1\bigl(\mathrm{tr}\bigl[ \Phi\Phi^\dagger \bigr]
  \bigr)^2 - g_2\mathrm{tr}\bigl[ \Phi\Phi^\dagger \Phi\Phi^\dagger
  \bigr]
 \end{split}
\label{eq:Lag_u2u2}
\end{equation}
with a complex $2\times2$ matrix $\Phi$.  This Lagrangian density is
invariant under the transformation,
\begin{equation}
 \Phi \longrightarrow V_L\Phi V_R^\dagger \;,
\end{equation}
where $V_L$ and $V_R^\dagger$ are independent $\mathrm{U}(2)$
matrices.  In the context of chiral symmetry of QCD, the matrix $\Phi$
has a parametrization in terms of the hadronic degrees of freedom:
\begin{equation}
 \Phi = \Sigma + i \Pi = \sum^3_{a=0} t_a (\sigma_a + i\pi_a) \;,
\end{equation}
where the $t_a$ are the $\mathrm{u}(2)$ generators (and unity)
normalized according to $\mathrm{tr}[t_a t_b]=\delta_{ab}$.  We can
write
\begin{align}
 \Sigma &= \left(
  \begin{array}{cc}
   \frac{1}{\sqrt{2}} a^0 + \frac{1}{\sqrt{2}} \sigma & a^+ \\
   a^- & -\frac{1}{\sqrt{2}} a^0 + \frac{1}{\sqrt{2}} \sigma
  \end{array} \right) \;, \\
 \Pi &= \left(
  \begin{array}{cc}
   \frac{1}{\sqrt{2}} \pi^0 + \frac{1}{\sqrt{2}} \eta & \pi^+ \\
   \pi^- & -\frac{1}{\sqrt{2}} \pi^0 + \frac{1}{\sqrt{2}} \eta
  \end{array} \right) \;,
\end{align}
using the conventional notation in hadron physics; $a^0=\sigma_3$,
$a^\pm=(a^1\mp i a^2)/\sqrt{2}=(\sigma_1\mp i\sigma_2)/\sqrt{2}$,
$\sigma=\sigma_0$ for the scalar sector and $\pi^0=\pi_3$,
$\pi^\pm=(\pi^1\mp i\pi^2)/\sqrt{2}=(\pi_1\mp i\pi_2)/\sqrt{2}$,
$\eta=\pi_0$ for the pseudo-scalar sector.  [This $\eta$ does not
  correspond to the physical $\eta$ in nature but to the
  flavor-singlet $\eta_0$ which mixes partially with $\eta_3$ to
  become the physical $\eta'$.]  With this notation the Lagrangian
density in Euclidean space-time is expressed as
\begin{equation}
 \mathcal{L} = \partial_\mu\sigma\, \partial_\mu\sigma
  + \partial_\mu\vec{\pi}\cdot \partial_\mu\vec{\pi}
  + \partial_\mu\eta\, \partial_\mu\eta
  + \partial_\mu\vec{a}\cdot \partial_\mu\vec{a} + U_\Lambda \;,
\end{equation}
where we find the potential,
\begin{equation}
 \begin{split}
  U_\Lambda &= \frac{1}{2} \mu_\Lambda^2
   (\sigma^2 + \vec{\pi}^2 + \eta^2 + \vec{a}^2 ) \\
  &\quad + \Bigl( g_1 + \frac{1}{2}g_2 \Bigr)
   (\sigma^2 + \vec{\pi}^2 + \eta^2 + \vec{a}^2)^2 \\
  &\quad + 2g_2 \Bigl[ (\sigma^2 + \vec{\pi}^2)
   (\eta^2 + \vec{a}^2) - (\sigma\eta + \vec{\pi}\cdot\vec{a})^2
   \Bigr] \;,
 \end{split}
\label{eq:U}
\end{equation}
from Eq.~\eqref{eq:Lag_u2u2}.  To simplify the notation in what
follows, we introduce the new couplings and variables;
\begin{align}
  \lambda_1 &\equiv 4! \, \Bigl(g_1 + \frac{1}{2}g_2 \Bigr) \;,\qquad
  \lambda_2 \equiv 2g_2 \;,
\label{eq:lam12} \\
 \begin{split}
  \varphi &\equiv \sigma^2 + \vec{\pi}^2 + \eta^2 + \vec{a}^2 \;, \\
  \xi &\equiv (\sigma^2 + \vec{\pi}^2)(\eta^2 + \vec{a}^2)
   - (\sigma\eta - \vec{\pi} \cdot \vec{a})^2 \;.
 \end{split}
\label{eq:phi-xi}
\end{align}
Roughly speaking $\sigma$ plays the role of the component of $\phi$
that acquires a finite expectation value in the $\mathrm{O}(N)$ scalar
theory and $\varphi$ defined above is the counterpart of $\phi^2$.
Using these variables we can rewrite Eq.~\eqref{eq:U} into a concise
representation as follows:
\begin{equation}
 U_\Lambda(\varphi,\xi) = \frac{1}{2}\mu_\Lambda^2 \varphi
  + \frac{1}{4!}\lambda_{1\Lambda} \varphi^2
  + \lambda_{2\Lambda} \,\xi \;.
 \label{eq:tree_u2u2}
\end{equation}


\subsection{Functional Renormalization Group Equation}
\label{sec:FRG_u2u2}

The functional RG equation in this system with $\utwo$ symmetry has
the same form as the generic one~\eqref{eq:FRG} in the previous
section.  Thus, the functional RG equation in the LPA takes the
following form (with $d=3$):
\begin{equation}
 \frac{\partial U_k(\varphi,\xi)}{\partial k}
  = K_d\, k^{d+1} \sum_i \frac{1}{E^2_i} \;,
 \label{eq:FRG_u2u2}
\end{equation}
where the energies in the denominator are for $i=\sigma$, $\vec{\pi}$,
$\eta$, and $\vec{a}$, given with the mass eigenvalues obtained from
the potential curvature.  One might imagine that this straightforward
procedure should work if $U_k$ is given as an explicit function in
terms of $\sigma$, $\vec{\pi}$, $\eta$, and $\vec{a}$.  In fact,
however, the $\mathrm{U}(2)\times\mathrm{U}(2)$ symmetry constrains
the potential and $U_k$ is a function of only two independent
variables instead of four; that is, $U_k$ is parametrized by $\varphi$
and $\xi$ only as exemplified in Eq.~\eqref{eq:tree_u2u2}.

Below we will elucidate how to read off the expressions for the mass
eigenvalues from the derivatives of $U_k(\varphi,\xi)$, which requires
a particular strategy for the calculation.  Actually taking the second
derivative on $U_k(\varphi,\xi)$ is just a simple manipulation.  The
most non-trivial part lies in the following:  The resulting masses are
easily expressed in terms of $\sigma$, $\vec{\pi}$, $\eta$, and
$\vec{a}$, which must be converted into $\varphi$ and $\xi$ in the
end.

We can simplify the problem a bit by observing that one field
($\sigma$ in our case) acquires a non-vanishing expectation value and
others have a vanishing expectation value and can be set to zero in
the end.  This means that we should take $\varphi\to\sigma^2$ and
$\xi\to0$ after writing down the functional RG equations.  It is
therefore a good truncation to keep the $\xi$-dependent term only up
to the first few orders \cite{Berges:1996ja}:
\begin{equation}
 U_k(\varphi,\xi) = V_k(\varphi) + W_k(\varphi)\, \xi + O(\xi^2) \;.
\end{equation}
In the present case we truncate the above expansion up to the linear
order in $\xi$ (where $\xi$ is by definition of quartic order with
respect to the meson field variables), so that our results cover
at a minimum the same behavior that is captured in the perturbative
$\varepsilon$-expansion.  In other words, we can say that we have
chosen an initial condition where the coefficient of the $\xi^2$-order
term is zero (as in Eq.~\eqref{eq:tree_u2u2}), and a truncation where
it is set to remain zero throughout in the RG flow.  This is the
same-order truncation as the one used in
\cite{Berges:1996ja,Berges:1996ib}.

Let us demonstrate our strategy by taking an example of calculating
$M_\sigma^2$.  The second derivative immediately leads to
\begin{align}
 M_\sigma^2
  &= \frac{\partial^2 (V_k+W_k\xi)}{\partial\, \sigma^2} \notag\\
  &= 4\sigma^2(V_k''+W_k''\xi) + 2(V_k'+W_k'\xi) +2W_k\vec{a}^2 \notag\\
  &\qquad +8\sigma W_k' \bigl[ \sigma(\eta^2+\vec{a}^2) -
   \eta(\sigma\eta-\vec{\pi}\cdot\vec{a}) \bigr] \;.
\end{align}
It is highly non-trivial how to rewrite this expression for
$M_\sigma^2$ into a form in terms of $\varphi$ and $\xi$.  We will do
this by thinking of a particular situation where only $\sigma$ and
$a^1$ take a finite value.  This is an almost unique choice for
simplification;  we can take $\vec{\pi}$ and $\eta$ as small as we
like, but $\vec{a}$ is constrained by Eq.~\eqref{eq:phi-xi} for a
given set of $\varphi$ and $\xi$.  Then we have
\begin{equation}
 \varphi = \sigma^2 + (a^1)^2 \;,\qquad
 \xi = \sigma^2 (a^1)^2 \;,
\end{equation}
from which we can solve for $\sigma$ and $a^1$ as
\begin{equation}
 \sigma^2, (a^1)^2 \,=\,
  \frac{\varphi}{2} \pm \sqrt{\frac{\varphi^2}{4}-\xi} \;,
\label{eq:sigma-a1}
\end{equation}
which can be expanded in terms of $\xi$ into a form;
\begin{equation}
 \sigma^2 = \varphi - \frac{\xi}{\varphi} \;,\qquad
 (a^1)^2 = \frac{\xi}{\varphi} \;,
\end{equation}
up to the linear order of $\xi$ (higher-order terms are unnecessary in
the $\xi\to0$ limit).  These expanded forms look singular at
$\varphi\to0$, but one should keep in mind that the expansion assumes
$\varphi\gg\xi$ in Eq.~\eqref{eq:sigma-a1} and one can always choose
$\xi$ in such a way that $\xi$ is small enough to satisfy this
inequality.  Therefore the singularity at $\varphi\to0$ is only
spurious, as we will check explicitly later.

Substituting these expressions into $M_\sigma^2$ we find
\begin{equation}
 \begin{split}
  M_\sigma^2 &= 4V_k''\Bigl(\varphi-\frac{\xi}{\varphi}\Bigr)
   +4W_k''\varphi\,\xi \\
  &\qquad + 2V_k' + 10W_k'\xi + 2W_k\frac{\xi}{\varphi} \;.
 \end{split}
\end{equation}
In the same way we can read the mass expressions off for other meson
fields:
\begin{align}
 & M_{\pi^1}^2 = M_\eta^2
  = 2V_k' + 2W_k'\xi \;, \\
 & M_{\pi^2}^2 = M_{\pi^3}^2
  = M_{\pi^1}^2 + 2W_k\frac{\xi}{\varphi} \;, \\
 & M_{a^1}^2 = M_{\pi^1}^2 + 4V_k''\frac{\xi}{\varphi}
  + 8W_k'\xi + 2W_k\Bigl(\varphi-\frac{\xi}{\varphi}\Bigr) \;, \\
 & M_{a^2}^2 = M_{a^3}^2 = M_{\pi^1}^2
  + 2W_k \Bigl(\varphi-\frac{\xi}{\varphi}\Bigr) \;.
\end{align}
We remark here that, if $U_k(\varphi,\xi)$ contains a term of order
$\xi^2$, the mass $M_{a^1}^2$ would have an extra term proportional to
$\varphi\xi$, which is vanishing in the present case due to our choice
of the initial condition and the truncation.

It is important to note that the mass matrix does not only have
diagonal components but also off-diagonal ones which lead to a mixing
between different fields.  These off-diagonal components are given by
\begin{align}
 M_{\sigma\text{-}a^1}^2 &= 4(V_k'' + W_k + W_k'\varphi ) \sqrt{\xi} \;, \\
 M_{\eta\text{-}\pi^1}^2 &= 2W_k \sqrt{\xi} \;.
\end{align}
In order to make use of the simple expression for the functional RG
equation in Eq.~\eqref{eq:FRG_u2u2}, we have to identify the
eigenmodes $(\tilde{\sigma},\tilde{a}^1)$ for the $\sigma$-$a^1$ part
and $(\tilde{\eta},\tilde{\pi}^1)$ for the $\eta$-$\pi^1$ part.  It is
not difficult to find the eigenvalues of the $2\times2$ matrix spanned
in $\sigma$-$a^1$ space and the results are found to be
\begin{align}
 M_{\tilde{\sigma}}^2 &= 4V_k''\Bigl(\varphi-\frac{\xi}{\varphi}\Bigr)
  + 4W_k''\varphi\xi + 2V_k' + 10W_k'\xi \notag \\
 &\qquad + 2W_k\frac{\xi}{\varphi}
  + \frac{8(V_k''+W_k+W_k'\varphi)^2}{2V_k''-W_k}\frac{\xi}{\varphi}
  \;, \\
 M_{\tilde{a}^1}^2 &= 4V_k''\frac{\xi}{\varphi} + 2V_k' + 10W_k'\xi
  + 2W_k\Bigl(\varphi-\frac{\xi}{\varphi}\Bigr) \notag \\
 &\qquad -\frac{8(V_k''+W_k+W_k'\varphi)^2}{2V_k''-W_k}
  \frac{\xi}{\varphi} \;.
\end{align} 
In the same way, regarding the $\eta$-$\pi^1$ mixing, we diagonalize
the mass matrix, yielding
\begin{align}
 M_{\tilde{\eta}}^2 &= 2V_k' + 2W_k'\xi + 2W_k \sqrt{\xi} \;, \\
 M_{\tilde{\pi}^1}^2 &= 2V_k' + 2W_k'\xi - 2W_k \sqrt{\xi} \;.
\end{align} 

The remaining task is to decompose the functional RG equation into the
one contributing to $V_k(\varphi)$ and the other contributing to
$W_k(\varphi)$.  As seen from the LHS of Eq.~\eqref{eq:FRG_u2u2}, the
part contributing to $W_k(\varphi)$ must come from terms proportional
to $\xi$.  (In the above the terms of $O(\sqrt{\xi})$ cancel between
$M_{\tilde{\eta}}^2$ and $M_{\tilde{\pi}^1}^2$.)  Thus, we should
expand the RHS of Eq.~\eqref{eq:FRG_u2u2} and identify the flow
equations for $V_k(\varphi)$ and $W_k(\varphi)$ from the contributions
of $O(\xi^0)$ and $O(\xi^1)$, respectively.

The $O(\xi^0)$ contribution is easy to derive from
Eq.~\eqref{eq:FRG_u2u2}, that is,
\begin{equation}
 \frac{\partial V_k}{\partial k} = K_d\, k^{d+1}
  \Bigl( \frac{1}{E^2_\sigma} + \frac{4}{E^2_{\pi\eta}}
   + \frac{3}{E^2_a} \Bigr) \;,
\label{eq:FRG_U}
\end{equation}
where we define
\begin{align}
 & E_\sigma^2 = k^2 + 2V_k' + 4V_k''\varphi \;,\\
 & E_{\pi\eta}^2 = k^2 + 2V_k' \;,\\
 & E_a^2 = k^2 + 2V_k' + 2W_k\varphi \;.
\label{eq:masses}
\end{align}
The calculations in the order $O(\xi^1)$ are rather complicated.
After tedious but straightforward computations we finally obtain
\begin{align}
 \frac{\partial W_k}{\partial k} &= -K_d\, {k^{d+1}} \Biggl\{
  \frac{8W_k' + 4W_k\varphi^{-1}}{E_{\pi\eta}^4}
  -\frac{8W_k^2}{E_{\pi\eta}^6} \notag\\
 &\quad +\biggl[ -4V_k''\varphi^{-1} + 4W_k''\varphi + 10W_k' +
  2W_k\varphi^{-1} \notag\\
 &\quad\qquad +\frac{8(V_k''+ W_k + W_k'\varphi)^2}{2V_k''-W_k}
  \varphi^{-1} \biggr] \frac{1}{E_\sigma^4} \notag \\
 &\quad +\biggl[ 4V_k''\varphi^{-1} + 14W_k' - 6W_k\varphi^{-1}
  \notag \\
 &\quad\qquad -\frac{8(V_k'' + W_k + W_k'\varphi)^2}{2V_k''-W_k}
  \varphi^{-1}\biggr] \frac{1}{E_a^4} \Biggr\} \;.
\label{eq:FRG_V}
\end{align} 
The above equations~\eqref{eq:FRG_U}, \eqref{eq:masses}, and
\eqref{eq:FRG_V} are our central results at the algebraic level.  We
emphasize that, thanks to the choice of the optimized IR regulator,
these expressions no longer contain momentum integrals and are much
simpler compared to those in earlier works.  This also makes the
connection to perturbative RG results more apparent.

Now let us make sure that Eq.~\eqref{eq:FRG_V} is not singular at
$\varphi\to0$.  We see that the energy denominator becomes the same in
all terms, i.e.\ $E_\sigma^2 = E_{\pi\eta}^2 = E_a^2$, as the limit
$\varphi\to0$ is taken.  Then, we pick up the singular terms from the
parentheses in the RHS of Eq.~\eqref{eq:FRG_V}, whose coefficients
amount to
\begin{equation}
 \begin{split}
  & 4W_k + \biggl[ -4V_k'' + 2W_k
   + \frac{8(V_k''+W_k)^2}{2V_k''-W_k} \biggr] \\
  &\qquad\qquad + \biggl[ 4V_k'' - 6W_k - \frac{8(V_k''+W_k)^2}
   {2V_k''-W_k} \biggr] = 0 \;,
 \end{split}
\end{equation}
where the first term is due to the part with $E_{\pi\eta}^4$, the
second parenthesis to the part with $E_\sigma^4$, and the last
parenthesis to the part with $E_a^4$ in the functional RG
equation~\eqref{eq:FRG_V}.


\subsection{Flow in Parameter Space}

Let us check that our equations~\eqref{eq:FRG_U}, \eqref{eq:masses},
and \eqref{eq:FRG_V} are consistent with the known results from the
$\varepsilon$-expansion~\cite{Pisarski:1980ix,Pisarski:1983ms}.  To
this end we  expand the potential up to the fourth order or sixth
order in terms of the fields:
\begin{equation}
 V_k = \frac{1}{2}\mu_k^2 \varphi
  + \frac{1}{4!}\lambda_{1k} \varphi^2 + \zeta_{1k}\varphi^3 \;,
  \qquad
 W_k = \lambda_{2k} + \zeta_{2k}\varphi \;,
\label{eq:expanded}
\end{equation}
where a constant offset of the potential energy is dropped.  It should
be noted that $\xi$ is already of fourth order.  For notational
convenience, we will use $\lambda_{1k}$ and $\lambda_{2k}$ and will
change to the couplings $g_1$ and $g_2$ only at the end to make a
comparison with the $\varepsilon$-expansion results.

By expanding the flow equation in terms of $\varphi$, we can formulate
the flow of the coupling constants (where we will set
  $\zeta_{1k}=\zeta_{2k}=0$ and work up to the fourth order for the
moment) as
\begin{align}
 \frac{\partial V_k}{\partial k}
  &= \frac{1}{2}\frac{\partial \mu_k^2}{\partial k}\varphi
   + \frac{1}{4!}\frac{\partial \lambda_{1k}}{\partial k}\varphi^2
   \notag\\
  &= K_d\, k^{d+1} \Biggl( \frac{8}{E^2}
   - \frac{5\lambda_{1k} + 18\lambda_{2k}}{3E^4} \varphi \notag\\
  &\quad\qquad + \frac{ 4\lambda_{1k}^2 + 18\lambda_{1k}\lambda_{2k}
   + 108\lambda_{2k}^2 }{9 E^6} \varphi^2 \Biggr) \;,
\end{align}
where $E^2=k^2 + \mu_k^2$.  We also have
\begin{equation}
 \frac{\partial W_k}{\partial k}
 = \frac{\partial \lambda_{2k}}{\partial k}
 = K_d\, k^{d+1} \frac{8\lambda_{1k}\lambda_{2k} + 16\lambda_{2k}^2}
   {E^6} \;.
\end{equation}
We note that there is a cancellation which eliminates the terms
proportional to $\varphi^{-1}$ from the results, as we have already
checked before.  This allows us to easily decompose the flow equations
into
\begin{align}
 \frac{\partial \mu_k^2}{\partial k} &=
  -\frac{2K_d\, k^{d+1}}{3E^4} \bigl(5\lambda_{1k }+ 18\lambda_{2k}
  \bigr) \;,
\label{eq:flow_mu_u2u2} \\ 
 \frac{\partial \lambda_{1k}}{\partial k} &=
  \frac{16K_d\, k^{d+1}}{3E^6} \bigl( 2\lambda_{1k}^2 +
   9\lambda_{1k} \lambda_{2k} + 54\lambda_{2k}^2 \bigr) \;,
\label{eq:flow_lambda1_u2u2} \\ 
 \frac{\partial \lambda_{2k}}{\partial k} &=
  \frac{8K_d\, k^{d+1}}{E^6} \bigl(\lambda_{1k} \lambda_{2k}
  + 2\lambda_{2k}^2\bigr)  \;,
\label{eq:flow_lambda2_u2u2}
\end{align}
up to the fourth order for $d$ dimensions.  At this point, it is a
simple task to convert the above into the $\varepsilon$-expansion
results obtained by Pisarski and Wilczek \cite{Pisarski:1983ms}.  We
rescale the coupling constants in the following way (see also
Eq.~\eqref{eq:lam12}):
\begin{equation}
 \lambda_{1k} = 4!\,\frac{\pi^2}{3}\Bigl( \bar{g}_1
  + \frac{1}{2}\bar{g}_2 \Bigr) k^{4-d} \;,
 \quad
 \lambda_{2k} = \frac{2\pi^2}{3} \bar{g}_2 k^{4-d} \;,
\end{equation}
and use $\bar{\mu}_k^2=k^{-2}\mu_k^2$ as previously defined.  Then,
after a short calculation, in $d=4-\varepsilon$ dimensions, we readily
arrive at
\begin{align}
 k\frac{\partial \bar{g}_1}{\partial k}
  &= -\varepsilon \bar{g}_1 + \frac{8}{3}\bar{g}_1^2
  + \frac{8}{3}\bar{g}_1\bar{g}_2 + \bar{g}_2^2 \;,
\label{eq:flowbarg_1} \\
 k\frac{\partial \bar{g}_2}{\partial k}
  &= -\varepsilon \bar{g}_2 + 2\bar{g}_1\bar{g}_2
  + \frac{4}{3}\bar{g}_2^2 \;,
\label{eq:flowbarg_2}
\end{align}
up to quadratic order in $\varepsilon$ on the RHS, where we evaluated
$K_d$ at the expansion point $d=4$ and used $K_4=(32\pi^2)^{-1}$.
This clearly shows that our calculations correctly reproduce the
results in the $\varepsilon$-expansion in the leading order, as they
should.  One can prove that the flow with respect to $\bar{g}_1$ and
$\bar{g}_2$ (with $\bar{\mu}_k=0$) does not have any stable IR fixed
point~\cite{Pisarski:1983ms}.  This is a reasonable but indirect
argument that the phase transition should be a fluctuation-induced
first order one~\cite{Bak:1976zza}.  It is highly non-trivial why the
behavior of $\bar{g}_1$ and $\bar{g}_2$ is capable of describing a
form of the potential which admits a first-order phase transition.
Let us consider this question by taking account of the flow behavior
of $\bar{\mu}_k^2$, which already goes beyond the
$\varepsilon$-expansion (in which $\bar{\mu}_k^2=0$).

The set of equations~\eqref{eq:flow_mu_u2u2},
\eqref{eq:flow_lambda1_u2u2}, and \eqref{eq:flow_lambda2_u2u2} is our
convenient starting point to deal with the parameter flow because one
can intuitively connect $\lambda_{1k}$ and $\lambda_{2k}$ to the shape
of the potential;  $\lambda_{1k}$ is a coefficient of the quadratic
term $\varphi^2$ (i.e.\ the quartic term $\sigma^4$).  The flow
equations for the dimensionless couplings in $d=3$ dimensions,
$\bar{\mu}_k^2$, $\bar{\lambda}_{1k}$, and $\bar{\lambda}_{2k}$, are
\begin{align}
 & k\frac{\partial\bar{\mu}_k^2}{\partial k}
  = -2\bar{\mu}_k^2
    - \frac{1}{9\pi^2 (1+\bar{\mu}_k^2)^2}
   \bigl( 5\bar{\lambda}_{1k}+18\bar{\lambda}_{2k} \bigr) \;,
\label{eq:flow_mu_dl_u2u2} \\
 & k\frac{\partial\bar{\lambda}_{1k}}{\partial k}
  = -\bar{\lambda}_{1k} \notag\\
  &\qquad + \frac{8}{9\pi^2 (1+\bar{\mu}_k^2)^3}
   \bigl( 2\bar{\lambda}_{1k}^2
   + 9\bar{\lambda}_{1k} \bar{\lambda}_{2k}
   + 54\bar{\lambda}_{2k}^2 \bigr) \;,
\label{eq:flow_lambda1_dl} \\
 & k\frac{\partial\bar{\lambda}_{2k}}{\partial k}
  = -\bar{\lambda}_{2k}
  + \frac{4}{3\pi^2 (1+\bar{\mu}_k^2)^3}
  \bigl(\bar{\lambda}_{1k} \bar{\lambda}_{2k}
  + 2\bar{\lambda}_{2k}^2 \bigr) \;.
\label{eq:flow_lambda2_dl}
\end{align}
In the same way it is tedious but straightforward to write down the
differential equations including sixth-order couplings $\zeta_{1k}$
and $\zeta_{2k}$.  The expressions are easily derived from
Eqs.~\eqref{eq:FRG_U} and \eqref{eq:FRG_V} and they are too lengthy
to display here, so we shall give their explicit forms in the
  Appendix.

We solve these differential equations numerically.  The RHS of these
equations has no explicit dependence on the scale $k$ and the LHS can
be rewritten as $k\partial/\partial k=\partial/\partial t$ using
$t\equiv \ln(k/\Lambda)$, so that we can describe the RG flow in terms
of this variable.  We numerically trace the running of these couplings
from $t=0$ to $t=-2.3$ (i.e.\ $k=0.1\Lambda$) to determine the flow
diagram, using the fourth-order Runge-Kutta method.  We used an
adaptive step size ($<10^{-4}$) so that no numerical instability
occurs.  [In later discussions we will use the fifth-order
  Runge-Kutta-Fehlberg (4,5) (RKF45) method to solve the functional
  equations.  In the present case, where the partial differential
  equations are relatively simple, the fourth-order method is
  sufficient.]
\vspace{3mm}


\paragraph*{The case with $\bar{\lambda}_{2\Lambda}=0$
  (Fig.~\ref{fig:o8}):}
Let us begin with a simple case.  We see that $\bar{\lambda}_{2k}=0$
is a solution of Eq.~\eqref{eq:flow_lambda2_dl}.  If we start the RG
flow with the initial condition $\bar{\lambda}_{2\Lambda}=0$, this
means that $\bar{\lambda}_{2k}$ remains zero for any $k$ throughout
the flow and Eqs.~\eqref{eq:flow_mu_dl_u2u2} and
\eqref{eq:flow_lambda1_dl} determine the evolution of $\bar{\mu}_k^2$
and $\bar{\lambda}_{1k}$ with changing $k$ (up to the fourth-order of
the Taylor expansion).  In this case with $\bar{\lambda}_{2k}=0$ the
flow equations are reduced to those in the $\mathrm{O}(8)$-symmetric
scalar theory.  It is easy to make sure that we can also reproduce the
well-known results from the $\varepsilon$-expansion by setting
$\bar{\lambda}_{2k}=0$ and using $K_4=(32\pi^2)^{-1}$ in the above
expressions.  To go to the next-to-leading order in the
$\varepsilon$-expansion one must go beyond the local potential
approximation, as done in the derivative expansion
\cite{Berges:2000ew,Litim:2001dt} or in the so-called BMW
approximation \cite{Blaizot:2005xy}, which is beyond the scope of our
present approach.

Using Eqs.~\eqref{eq:flow_mu_dl_u2u2} and \eqref{eq:flow_lambda1_dl}
with $\bar{\lambda}_{2k}=0$, we plot the RG flow to obtain
Fig.~\ref{fig:o8}.  We find an IR fixed point at
\begin{equation}
 \begin{split}
 \bar{\mu}_\ast^2 &= -\frac{5}{37} \approx -0.135 \;,\\
 \bar{\lambda}_{1\ast} &= \frac{9\pi^2}{16}\Bigl(
  \frac{32}{37}\Bigr)^3 \approx 3.59 \;.
 \end{split}
\end{equation}
We find indeed in Fig.~\ref{fig:o8} a fixed point corresponding to the
second-order phase transition at these values of the couplings.  It is
obvious from Fig.~\ref{fig:o8} that there is no RG trajectory flowing
from the $(\bar{\mu}_k^2<0,\,\bar{\lambda}_{1k}>0)$ region into the
$(\bar{\mu}_k^2>0,\,\bar{\lambda}_{1k}<0)$ region.  Thus it is not
possible to arrive at a fluctuation-induced first-order phase
transition as long as the initial $\bar{\lambda}_{2\Lambda}$ is chosen
to be zero.  We see that such a flow which would favor a
fluctuation-induced first-order transition is prevented by the
presence of the UV fixed point at the origin from which the flow
emerges.
\vspace{3mm}


\paragraph*{The case with $\bar{\lambda}_{2\Lambda}\neq0$
  (Figs.~\ref{fig:region04-4}, \ref{fig:region04-6},
  \ref{fig:region07-6}):}
Now we proceed to the case with a finite $\bar{\lambda}_{2k}$.
Interestingly enough, the flow pattern changes drastically once
$\bar{\lambda}_{2\Lambda}$ deviates from zero.  Then we have to
analyze the flow pattern as a function of $\bar{\mu}_k^2$,
$\bar{\lambda}_{1k}$, and $\bar{\lambda}_{2k}$.  To visualize the flow
pattern, we solve Eqs.~\eqref{eq:flow_mu_dl_u2u2},
\eqref{eq:flow_lambda1_dl}, and \eqref{eq:flow_lambda2_dl} for a
particular initial condition with a certain choice of
$\bar{\lambda}_{2\Lambda}$, and investigate the destination of the
flow in the $\bar{\mu}_k^2$-$\bar{\lambda}_{1k}$ plane.  As we have
already elucidated in Fig.~\ref{fig:schematic} in the $\mathrm{O}(8)$
case and also in Figs.~\ref{fig:region04-4}, \ref{fig:region04-6}, and
\ref{fig:region07-6}, we can classify the parameter regions according
to qualitative features of the potential form;
\begin{center}
 \begin{tabular}{l@{\hspace{1em}}c@{\hspace{1em}}l}
  $\bar{\mu}_k^2<0$ and $\bar{\lambda}_{1k}>0$ & $\to$ &
   Symmetry-broken \\
  $\bar{\mu}_k^2>0$ and $\bar{\lambda}_{1k}>0$ & $\to$ &
   Symmetric \\
  $\bar{\mu}_k^2>0$ and $\bar{\lambda}_{1k}<0$ & $\to$ &
   Necessary for Double-well
 \end{tabular}
\end{center}
as $k\to0$.

We have already discussed our results in Sec.~\ref{sec:results}.  As a
final remark in this subsection, we point out an interesting
consequence from the flow equations.

From an analysis of the flow equations~\eqref{eq:flow_mu_dl_u2u2},
\eqref{eq:flow_lambda1_dl}, and \eqref{eq:flow_lambda2_dl} we can
indeed confirm that in this approximation no fixed point with
$\lambda_{2k} \neq 0$ exists in $d=3$ dimensions, and in the presence
of a non-zero coupling $\lambda_{2k}$ a second-order phase transition
associated with this fixed point cannot be accommodated.

To see this, the fixed point
$(\bar{\mu}_\ast^2,\bar{\lambda}_{1\ast},\bar{\lambda}_{2\ast})$ is
located by imposing the condition that the RHS's of
Eqs.~\eqref{eq:flow_mu_dl_u2u2}, \eqref{eq:flow_lambda1_dl}, and
\eqref{eq:flow_lambda2_dl} vanish.  For
$\bar{\lambda}_{2\ast}\neq 0$, the fixed point condition can be
simplified and the resulting system of three equations for the three
couplings $\bar{\mu}_\ast^2$, $\bar{\lambda}_{1\ast}$, and
$\bar{\lambda}_{2\ast}$ can be solved analytically.  We find that for
$d=3$ no solutions for real values of the couplings exist (that is,
$\bar{\mu}_\ast^2$, $\bar{\lambda}_{1\ast}$, and
$\bar{\lambda}_{2\ast}$ are all complex then), and that hence the RG
flow in this approximation does not admit a second-order phase
transition.


\subsection{Functional Solution}

Without an evaluation of the higher-order terms we cannot locate
exactly where the fluctuation-induced first-order phase transition
takes place on Figs.~\ref{fig:region04-4}, \ref{fig:region04-6}, and
\ref{fig:region07-6}.  Nevertheless, it is clear that the second-order
critical line is overridden by the region which satisfies the
condition for a double-well potential as the initial value for
$\lambda_{2\Lambda}$ grows (see particularly Figs.~\ref{fig:o8} and
\ref{fig:region04-4}).

We can check if the analytical study in the previous subsection is
qualitatively correct by looking at the full functional solution
obtained with the grid method.  Since there is no scaling property
expected in the case of the fluctuation-induced first-order phase
transition, we plot the results in terms of unscaled variables (made
dimensionless not with $k$ but by the UV scale $\Lambda$).  We give
the numerical values for $\mu_k^2$, $\lambda_{1k}$, and $\lambda_{2k}$
in units of $\Lambda$ and we omit $\Lambda$ hereafter.


\begin{figure}
 \includegraphics[width=\columnwidth]{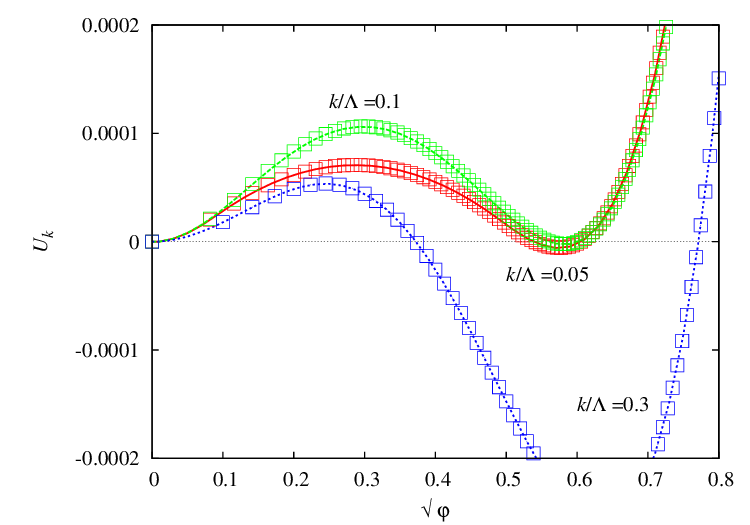}
 \caption{Evolution of the form of the potential with decreasing
   $k/\Lambda$ in the case with $\lambda_{2\Lambda}=0.4$.  The solid
   curves represent the results from our grid method (see the text for
   details) and the squares represent the results obtained in the
   method formulated in \cite{Adams:1995cv}.  The potential is given
   as a function of $\sqrt{\varphi}$.  The initial parameters are
   chosen as $\mu_\Lambda^2=-0.05$ and $\lambda_{1\Lambda}=0.35$.}
\label{fig:poten_compare}
\end{figure}


In the numerical calculations, we used a simple formulation for the
grid method.  We first discretize the function $U_k(\varphi)$ in terms
of $\sigma=\sqrt{\varphi}$ instead of $\varphi$, since we are
interested in the characteristic form of the effective potential in
the vicinity of $\varphi\sim0$ or $\sigma\sim0$.  It is thus more
suitable to use $\sigma$ to attain more resolution in the
small-$\varphi$ region.  Our mesh size is $\Delta\sigma=10^{-2}$ and
the mesh spans the interval between $\sigma=\pm2$.  The
  singularity in the flow equations at $\sigma=0$ is only spurious due
  to cancellation as we have checked.  In the numerical calculation
  one should remove this singularity analytically or discretize the
  fields not to hit the origin.  We here took the latter, i.e.\ we
  choose the support points such that the origin is not included;
  $\sigma_i=\Delta\sigma(i+0.5)$ with $i$ ranging from
  $-200$ to $199$.
To evaluate the derivatives $U_k'$ and $U_k''$ we used the 7-point
formula.  At the boundaries, where we cannot take $7$ points
(i.e.\ $3$ adjacent sites from the point where the derivatives are
evaluated), we used the 5-point formula and the 3-point formula, and
eventually at the very end we approximately used the next-site value.
These procedures enhance numerical errors locally near the edges, but
if the edges are sufficiently far from the small-$\varphi$ region of
our interest, those errors do not influence the results.  The step
size $\Delta k$ is an adaptive variable so that the computation
proceeds without numerical instability.  Typically $\Delta k$ is less
than $10^{-4}$.  Then we found no instability in the fourth-order
Runge-Kutta method.  In summary, after discretization we treat the
function $U_k(\varphi)$ as an array labelled by site number $i$ and
then proceed as follows:

\begin{enumerate}
 \item Choose an initial value of $k=\Lambda$ and fix the potential
   parameters $\mu_\Lambda$, $\lambda_{1\Lambda}$, and
   $\lambda_{2\Lambda}$ which also determines the initial
   configuration of the array for $V_\Lambda(\sigma)$ and
   $W_\Lambda(\sigma)$ (where $\sigma=\sqrt{\varphi}$).

 \item Evaluate the RHS of Eqs.~\eqref{eq:FRG_U} and \eqref{eq:FRG_V}
   using $V_k(\sigma)$ and $W_k(\sigma)$, and the numerical
   derivatives.  Update the potential from $V_k(\sigma)$ and
   $W_k(\sigma)$ down to $V_{k-\Delta k}(\sigma)$ and
   $W_{k-\Delta k}(\sigma)$ with the Runge-Kutta method.

 \item Iterate the above calculation until the scale $k$ reaches zero
   or a sufficiently small value, so that the full effective
   potential $V_0(\sigma)$ and $W_0(\sigma)$ results at the end from
   the functional RG flow.
\end{enumerate}

Such a discretization scheme may look very simple compared to more
sophisticated processes such as the one proposed in
\cite{Adams:1995cv}, which has actually been used in the earlier works
\cite{Berges:1996ja,Berges:1996ib}.  In this method $U_k$ and $U'_k$
are solved for by the evolution equations and $U''_k$ and $U'''_k$ are
determined by the matching condition.  While the idea sounds very
different from the simple numerical derivatives that we used, the
matching condition leads in effect to a natural generalization of the
formula for the numerical differentiation.  The accuracy is likely
better than that of the the 7-point formula.  In addition, for the
method of \cite{Adams:1995cv}, usually the RKF45 method (fifth-order
Runge-Kutta method with the fourth-order embedded for the error
control) is used.  This sophisticated method is very powerful, but at
the same time, it is difficult to apply it to solve a flow equation
with many terms, such as Eq.~\eqref{eq:FRG_V}.

We prefer to use the former simpler method as long as it works and we
do not need high precision data.  In fact we have solved the RG
equations making use of both the simple and sophisticated method, and
we present a quantitative comparison in Fig.~\ref{fig:poten_compare}.
We clearly see that the potential exhibits a double-well form as a
result of fluctuations and the deviation by two methods is not visible
in the figure.  We can conclude that our method is good enough for our
purpose.

We have already discussed our central results in
Sec.~\ref{sec:results} and so we do not reiterate the discussion here.
For the rest of this subsection we mention on some notable features in
the potential evolution.


\begin{figure}
 \includegraphics[width=\columnwidth]{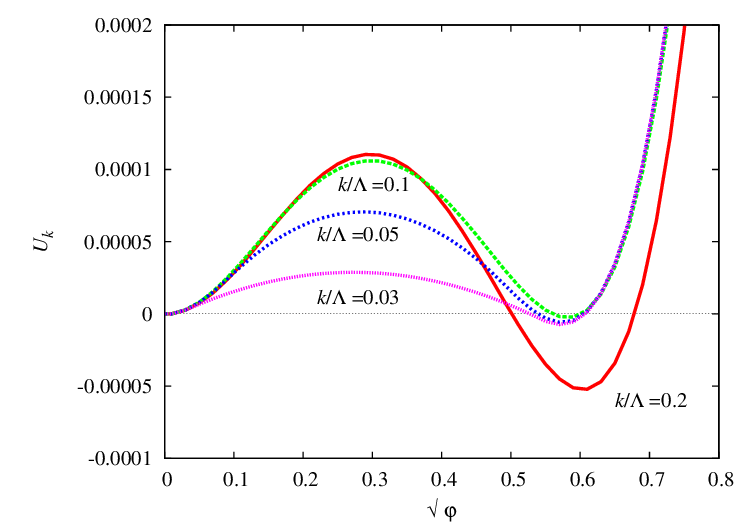}
 \caption{Evolution of the form of the potential in the region with
   $k/\Lambda<0.2$ in the case with $\lambda_{2\Lambda}=0.4$ as
   $k/\Lambda$ decreases.  At $k/\Lambda=1$ the initial parameters are
   chosen as $\mu_\Lambda^2=-0.05$ and $\lambda_{1\Lambda}=0.35$.}
\label{fig:poten_magnify}
\end{figure}


Figures~\ref{fig:poten_compare} and \ref{fig:poten_magnify} show the
evolution of the shape of the potential in the case of
$\lambda_{2\Lambda}=0.4$ starting with the initial condition
$\mu_\Lambda^2=-0.05$ and $\lambda_{1\Lambda}=0.35$, which is almost
at the first-order phase transition point (as indicated by the solid
line in Figs.~\ref{fig:region04-4} and \ref{fig:region04-6}).  We
stopped the evolution at $k/\Lambda=0.1$, which is legitimate because
the effective potential becomes convex when $k/\Lambda$ goes to zero
and the double-well shape of the potential is less manifest
\cite{Ringwald:1989dz}.  Here this happens below $k/\Lambda\sim 0.1$,
as shown in Fig.~\ref{fig:poten_magnify}.  We can clearly see in
Fig.~\ref{fig:poten_magnify} that the location of the minima does not
change for $k/\Lambda<0.1$, although the bump of the potential is
becoming flat with smaller $k/\Lambda$.  This is completely consistent
with the observation in Ref.~\cite{Berges:1996ja} and justifies our
prescription to stop the RG evolution at $k/\Lambda=0.1$ in order to
judge if the potential has a double-well shape.  Even though we choose
the step size $\Delta k$ as an adaptive variable, moreover, our
algorithm loses stability at $k/\Lambda\sim 0.02$ and cannot reach the
completely flat shape.  We observed the same behavior even using the
matching method with the RKF45 method.  Probably the implicit method
(i.e.\ Lax method) might cure this problem.  In any case, since our
aim is simply to determine the transition line on the phase diagram
shown in Figs.~\ref{fig:region04-4} and \ref{fig:region04-6}, there is
no practical need to require the potential form at $k/\Lambda<0.1$.


\begin{figure}
 \includegraphics[width=\columnwidth]{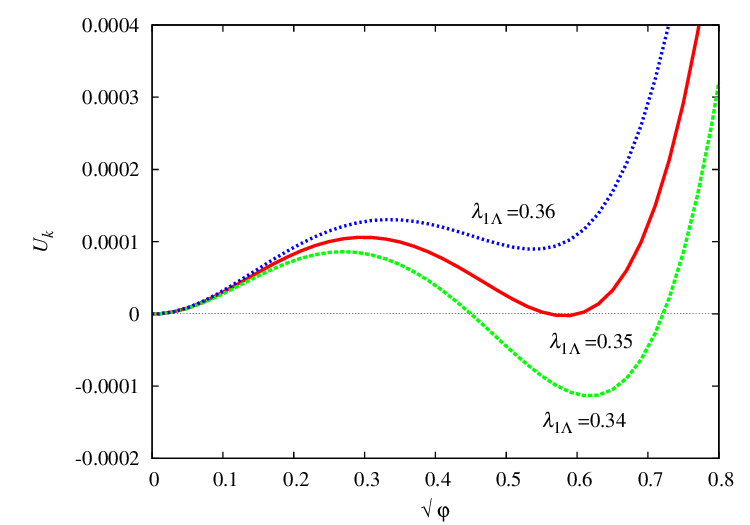}
 \caption{Change of the form of the potential as the initial value
   $\lambda_{1\Lambda}$ changes around $0.35$.  All results for
   the potential are given at fixed $k/\Lambda=0.1$.  Other initial
   parameters are all chosen to be the same as in
   Fig.~\ref{fig:poten_compare}.}
\label{fig:poten2}
\end{figure}


The first-order phase transition occurs with varying $\mu_\Lambda^2$
and/or $\lambda_{1\Lambda}$ for a given $\lambda_{2\Lambda}$.  Let us
fix $\mu_\Lambda^2$ at $-0.05$ here and change $\lambda_{1\Lambda}$ to
see how the first-order transition arises.  Our results are depicted
in Fig.~\ref{fig:poten2}.  Again, we have stopped the evolution at
$k/\Lambda=0.1$, which is sufficient to find a first-order phase
transition.  It is obvious that there is a peculiar change of the
shape of the potential which is associated with the first-order phase
transition from $\lambda_{1\Lambda}=0.34$ to $0.36$.  More precisely,
we can locate the transition point by detecting the second minimum
height and using the midpoint method for $\lambda_{1\Lambda}$, which
leads to a critical value of $\lambda_{1\Lambda}=0.3509$.  We picked
just one example of particular values of $\mu_\Lambda^2$ but can
repeat the same procedure with different parameters to find a
first-order phase transition line as indicated in
Figs.~\ref{fig:region04-4}, \ref{fig:region04-6}, and
\ref{fig:region07-6}.


\begin{figure}
 \includegraphics[width=\columnwidth]{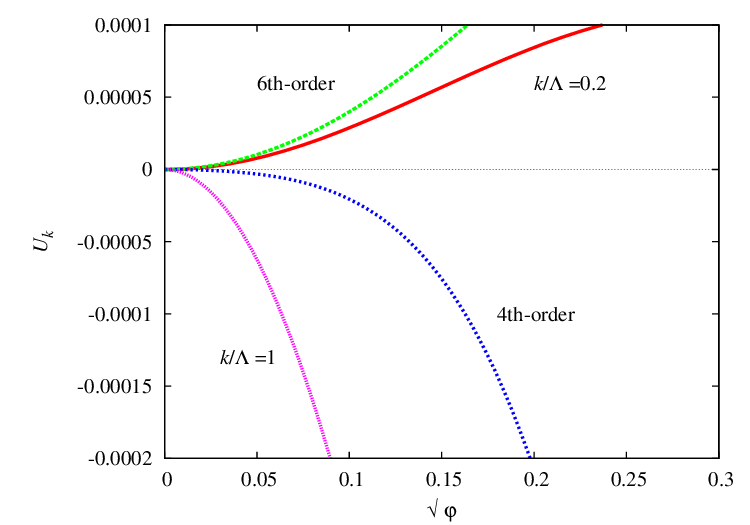}
 \caption{Comparison between the full effective potentials at
   $k/\Lambda=0.2$ and the expanded results from the Taylor expansion
   up to fourth order and sixth order.  The effective potential at
   $k/\Lambda=1$ is presented for reference.}
\label{fig:poten_exp}
\end{figure}


Finally let us discuss the comparison between the full functional
solutions and the Taylor expansion results.  We present
Fig.~\ref{fig:poten_exp} to show the expanded results according to
Eq.~\eqref{eq:expanded} together with the outputs from the grid
calculation.  It may be surprising at first glance that the curvature
at $\sqrt{\varphi}\simeq0$ has such large deviations from the
fourth-order results and the sixth-order or full grid results.  We
observe that the agreement in the curvature stays good as long as
$k/\Lambda$ is not too small ($k/\Lambda\sim 0.5$ for instance), for
which the curvature is still negative.  Even though the Taylor
expansion in terms of the field $\varphi$ should in principle work
well in the small-$\varphi$ region, the fourth-order expansion is not
good enough to capture the first-order phase transition
quantitatively.  We can see that the agreement in the curvature is
significantly improved by proceeding to the sixth-order expansion,
though the agreement in the quartic coefficient (i.e.\ $\lambda_{1k}$)
seems to be poor, which might be improved with inclusion of further
higher-order contributions.


\section{CONCLUSIONS}
\label{sec:conclusions}

We have applied the functional RG technique in the form of Wetterich's
flow equation to an analysis of a scalar theory with $\utwo$ symmetry.
We have been able to recover the perturbative results by means of
expansion in terms of the field, which also encompasses the standard
$\varepsilon$-expansion results \cite{Pisarski:1983ms}.  The choice of
a suitable regulator function and the resulting simple form of the RG
flow equations make this connection apparent.  Going beyond a
perturbative expansion in small couplings, we have further used the
functional RG flow equation in the local potential approximation to
study the emergence of a first-order phase transition.  To this end,
we have discretized the effective potential on a mesh grid to obtain
the global shape of the potential for a set of choices for the initial
conditions.  Our results confirm those of earlier studies
\cite{Berges:1996ja,Berges:1996ib}.

In the $\utwo$ scalar theory there are two fourth-order couplings --
$\lambda_{1k}$ appearing in the quartic coefficient in the effective
potential and $\lambda_{2k}$ which has no counterpart in the
$\mathrm{O}(N)$ theory -- in addition to the curvature $\mu_k^2$.  We
have found that the second-order critical line in the
$\bar{\mu}_k^2$-$\bar{\lambda}_{1k}$ plane is gradually overridden by
the first-order phase transition as the initial value
$\bar{\lambda}_{2\Lambda}$ starts to differ from zero.  We find
analytically that no fixed point associated with a second-order
transition exists for $d=3$ and $\bar{\lambda}_{2k}\neq 0$ in our
approximation which is perturbative but beyond the
$\varepsilon$-expansion.

In view of the expanded potential, a double-well shape of the
potential is possible in the region $\bar{\mu}_\Lambda^2>0$ and
$\bar{\lambda}_{1\Lambda}<0$, which would admit a first-order
transition if the potential is stabilized by higher-order terms,
which we investigated up to the sixth-order expansion.  We note that
a similar analysis has been performed in Ref.~\cite{Litim:1994jd} up
to the eighth-order expansion for the Coleman-Weinberg potential and
our results are qualitatively consistent with those of
Ref.~\cite{Litim:1994jd}.  From the flow diagram we learn that one way
to understand the fluctuation-induced first-order phase transition is
that this first-order phase transition at the tree level with
$\bar{\mu}_\Lambda^2>0$ and $\bar{\lambda}_{1\Lambda}<0$ penetrates
into other (physical) parameter regions for
$\bar{\lambda}_{2\Lambda}\neq 0$.  This is an interesting point of
view and gives an intuitive insight into the emergence of a
fluctuation-induced first-order phase transition.

Regarding the application to QCD physics, it would be intriguing to
study the chiral phase transition with both the meson
fluctuations~\cite{Schaefer:2004en,Nakano:2009ps} and the effective
restoration of $\UA$ symmetry taken into account.  So far, for
example, the so-called Columbia diagram in the mass plane of light and
heavy quarks has been investigated as a function of the $\UA$-breaking
strength~\cite{Fukushima:2003fw,Fu:2007xc}.  The possibility of the
fluctuation-induced first-order phase transition, however, has not yet
been considered because the model study was at the mean-field level.
The present work gives a theoretical framework necessary for such
investigations in the future.


\acknowledgments
We thank Jean-Paul Blaizot, Jan Pawlowski, and Jens Braun for useful
discussions.
K.~F.\ was supported by Grant-in-Aid for Young Scientists B
(No.\ 20740134) and also supported in part by Yukawa International
Program for Quark Hadron Sciences.
K.~K.\ was supported by Grant-in-Aid for JSPS Fellows (No.\ 22-3671).
B.~K.\ acknowledges support by the Research Cluster ``Structure and
Origin of the Universe''.


\appendix

\section{FLOW EQUATIONS IN THE SIXTH-ORDER TAYLOR EXPANSION}

Here we list a set of differential equations involving the sixth-order
Taylor coefficients $\zeta_1$ and $\zeta_2$.  We note that
Eq.~\eqref{eq:flow_mu_u2u2} for $\mu_k^2$ has no modification.
Equations~\eqref{eq:flow_lambda1_u2u2} and
\eqref{eq:flow_lambda2_u2u2} are replaced, respectively, by
\begin{align}
 \frac{\partial\lambda_{1k}}{\partial k} &=
  \frac{16K_d k^{d+1}}{3E^6} \bigl[ 2\lambda_{1k}^2
  +9\lambda_{1k}\lambda_{2k}+54\lambda_{2k}^2 \notag\\
  &\qquad\qquad\qquad\qquad\quad
  -27E^2(12\zeta_{1k}+\zeta_{2k}) \bigr] \;,\\
 \frac{\partial\lambda_{2k}}{\partial k} &=
  \frac{8K_d k^{d+1}}{E^6}\bigl( \lambda_{1k}\lambda_{2k}
  +2\lambda_{2k}^2 -4\zeta_{2k}E^2 \bigr) \;,
\end{align}
which are reduced to Eqs.~\eqref{eq:flow_lambda1_u2u2} and
\eqref{eq:flow_lambda2_u2u2} for $\zeta_{1k}=\zeta_{2k}=0$.  The
differential equations for $\zeta_{1k}$ and $\zeta_{2k}$ are
\begin{widetext}
\begin{align}
 \frac{\partial\zeta_{1k}}{\partial k} &=
  -\frac{K_d k^{d+1}}{108E^8}\bigl\{
  17\lambda_{1k}^3+54\lambda_{1k}^2\lambda_{2k}
  +648\lambda_{1k}\lambda_{2k}^2+2592\lambda_{2k}^3
  -216E^2\bigl[ 2\zeta_{1k}(11\lambda_{1k}+18\lambda_{2k})
   +\zeta_{2k}(\lambda_{1k}+12\lambda_{2k}) \bigr] \bigr\} \;,\\
 \frac{\partial\zeta_{2k}}{\partial k} &=
  -\frac{4K_d k^{d+1}}{3E^8} \bigl\{
  3\lambda_{2k}(2\lambda_{1k}^2+11\lambda_{1k}\lambda_{2k}
  +6\lambda_{2k}^2) - E^2 \bigl[ 432\zeta_{1k}\lambda_{2k}
  +\zeta_{2k}(23\lambda_{1k}+114\lambda_{2k}) \bigr] \bigr\} \;,
\end{align}
\end{widetext}
which were used to obtain the sixth-order results shown in
Figs.~\ref{fig:region04-6}, \ref{fig:region07-6}, and
\ref{fig:poten_exp}.


\end{document}